\documentclass[%
 reprint,
nofootinbib,
 amsmath,amssymb,
 aps,
 prl,
floatfix,
]{revtex4-2}

\usepackage{graphicx}
\usepackage{dcolumn}
\usepackage{bm}

\usepackage{epsfig}
\usepackage{amsmath}
\input{epsf}
\usepackage{psfrag}
\usepackage{xcolor}

\newcommand{\bea}{\begin{eqnarray}}
\newcommand{\eea}{\end{eqnarray}}
\newcommand{\nn}{\nonumber}

\begin{document}

\title{The Nucleon Energy Correlators}

\author{Xiaohui Liu}
 \email{xiliu@bnu.edu.cn}
 \affiliation{Center of Advanced Quantum Studies, Department of Physics, Beijing Normal University, Beijing, 100875, China}
 \affiliation{Center for High Energy Physics, Peking University, Beijing 100871, China}
\author{Hua Xing Zhu}%
 \email{zhuhx@zju.edu.cn}
\affiliation{Zhejiang Institute of Modern Physics, Department of Physics, Zhejiang University, Hangzhou, 310027, China}%

\begin{abstract}
We introduce the concept of the nucleon energy correlators, a set of novel objects that encode the microscopic details of a nucleon, such as the parton angular distribution in a nucleon, the  collinear splitting to all orders, as well as the internal transverse dynamics of the nucleon. The nucleon energy correlators complement the conventional nucleon/nucleus tomography, but without introducing the non-perturbative fragmentation functions or the jet clustering algorithms. We demonstrate how the nucleon energy correlators can be measured in the lepton-nucleon deep inelastic scattering. The predicted distributions display a fascinating phase transition between the perturbative and non-perturbative regime. In the perturbative phase, a polar angle version of the Bjorken scaling behavior is predicted. We discuss its possible applications and expect it aggrandize the physics content at the electron ion colliders with a far-forward detector.   
\end{abstract}

\maketitle

\textbf{\emph{Introduction.}}
The femtoscale structure of the nucleon
has been the central scientific importance of nuclear physics for decades. The next generation QCD facilities~\cite{,AbdulKhalek:2021gbh, Proceedings:2020eah,Anderle:2021wcy} will boost the revelation of the nucleon/nucleus partonic structure in great detail. Conventional approach to the nucleon/nucleus tomography is to probe its transverse momentum dependent
(TMD) structure functions through either the semi-inclusive deep inelastic scattering (SIDIS)~\cite{Collins:2005ie,Vogelsang:2005cs,HERMES:2009lmz,Bacchetta:2011gx,Echevarria:2014xaa,Scimemi:2019cmh} or the jet-based studies~\cite{Gutierrez-Reyes:2018qez,Liu:2018trl,Gutierrez-Reyes:2019msa,Gutierrez-Reyes:2019vbx,Arratia:2020nxw,Liu:2020dct,Arratia:2020ssx,Li:2020rqj,Kang:2020fka,H1:2021wkz,Kang:2021kpt,Liu:2021ewb,Kang:2021ffh,Li:2021gjw,Lai:2022aly,Kang:2022dpx}. However SIDIS calls for the knowledge of the TMD fragmentation functions, while jets involve the clustering procedure and require high machine energy, either probes seem to complicate the analysis in one way or another. 

There are other substitutions to jets and identified hadrons, such as event-shape observables and the energy-energy correlator (EEC)~\cite{Basham:1978bw,Basham:1978zq}. Recently, there has been ongoing efforts to reformulate jet substructure physics using the EEC and its higher point generalization~\cite{Chen:2020vvp}. This is largely inspired by the unprecedented 
detector resolution at the LHC, as well as insights from conformal collider physics~\cite{Hofman:2008ar,Belitsky:2013ofa,Belitsky:2013xxa,Kologlu:2019mfz,Korchemsky:2019nzm,Dixon:2019uzg}. For recent application of energy correlators in jet substructure see for example~\cite{Chen:2019bpb,Chen:2020adz,Chang:2020qpj,Li:2021zcf,Jaarsma:2022kdd,Komiske:2022enw,Holguin:2022epo,Yan:2022cye,Chen:2022jhb,Chang:2022ryc,Chen:2022swd,Lee:2022ige,Larkoski:2022qlf,Ricci:2022htc,Yang:2022tgm}. 

The EEC measures the correlation $\langle {\cal E}(n_i) {\cal E}(n_j) \rangle$ between the energy deposit in two detectors along directions $n_i$ and $n_j$ with angular separation $\theta_{ij}$, where ${\cal E}(n) = \lim_{r\to \infty} \int_0^\infty dt T_{0 \vec{n}}(t, \vec{n} r ) r^2 $ is the asymptotic energy flow operator with $T_{\mu\nu}$ the energy-stress tensor~\cite{Sveshnikov:1995vi,Tkachov:1995kk,Korchemsky:1999kt,Bauer:2008dt}. 
One particular interesting feature of the EEC is its collinear limit as $\theta_{ij} \to 0$, in which the EEC exhibits a universal scaling behavior~(modulo running coupling effects),
$\lim_{n_j \to n_i} {\cal E}(n_i){\cal E}(n_j) \sim \theta^{\gamma(\alpha_s)}$ as described by the light-ray OPE~\cite{Hofman:2008ar,Kologlu:2019mfz}. The collinear limit is well encoded in the EEC jet function~\cite{Dixon:2019uzg}
\bea\label{eq:jet_eec} 
J^q_{EEC}(\theta^2)  
&=& \sum_X
\sum_{{i,j\in X}} \!\!\!
\frac{{{\bar n}\!\!\!/}_{\alpha\beta}}{2}
\langle \Omega | {\bar \chi}_n \, 
\delta_{Q,{\cal P}_n}
\delta(\theta^2 - \theta^2_{ij} )
|X 
\rangle_{\alpha}  \nn \\
&& 
\hspace{5.ex}
\times \frac{E_iE_j}{(Q/2)^2}
\langle X|\chi_n|\Omega \rangle_\beta\,,
\eea 
where $\chi_n$ is the gauge invariant $n$-collinear quark field in the soft collinear effective theory (SCET)~\cite{Bauer:2000yr,Bauer:2001yt,Bauer:2001ct,Beneke:2002ph,Bauer:2002nz}
 that serves the source to create collinear particles out of the vacuum $|\Omega \rangle$ and $Q$ represents the hard scale that initiates the process. 
 $E_i$ and $E_j$ are the energies measured. The light like vectors $n=(1,\vec{0}_\perp,1)$ and ${\bar n}=(1,\vec{0}_\perp,-1)$. When $\theta Q \gg \Lambda_{\rm QCD}$, the EEC jet function can be predicted using the collinear splitting functions~\cite{Dixon:2019uzg,Chen:2019bpb}. When $\theta Q \sim \Lambda_{\rm QCD}$, a striking confining transition was observed in analyzing CMS Open Data~\cite{Komiske:2022enw}.  
 
The EEC has also been adapted to the TMD studies in DIS, where the back-to-back limit $\theta_{ij} \to \pi$ is probed instead~\cite{Li:2020bub,Ali:2020ksn,Li:2021txc}. It was shown that when the EEC is measured in DIS in this limit, the unpolarized TMD parton distribution arises~\cite{Li:2020bub,Li:2021txc}. However, given that the EEC essentially measures the {\it{final state correlations}} and carries no information on the nucleon, the current EEC probe of the TMDs is indirect, in the sense that the EEC is used as a simple replacement of the jets or hadrons, while its power and features are not yet fully exploited.  

In this manuscript, we propose a novel energy correlator named the nucleon EEC that probes the initial-final state correlation. The nucleon EEC encodes the information on the nucleon three dimensional microscopic structures, meanwhile inherits  the fascinating features of the conventional final-state EEC. As one of the major results of this work, we will demonstrate the accessibility to the nucleon EEC via the measurement in DIS
\bea\label{eq:eec-def1} 
\Sigma_N(Q^2,\theta^2)   
= \sum_i \int d\sigma(x_B,Q^2,p_i) \, 
x_B^{N-1} \frac{{\bar n} \! \cdot \! p_i }{P}  \, \delta(\theta^2 - \theta_i^2  )
\,,\quad  
\eea
where $N \ge 1$ is a positive power and $d\sigma$ is the differential cross section. $x_B$ is the Bjorken variable and $\theta_i$ is the polar angle of the calorimeter measured with respect to the beam. $p_i$ denotes the momentum flow into the detector and $P$ the momentum of the incoming nucleon while $Q^2$ the virtuality of the photon. We leave the detailed explanation to the rest of this manuscript.

\textbf{\emph{The nucleon energy-energy-correlator.}} 
We first generalize the EEC jet function to
introduce the (unpolarized) nucleon EEC, whose definition is
 \bea\label{eq:beam_eec}
 f^q_{EEC}(N,\theta^2)    
&=& \sum_{X}\, \sum_{{ k\in X} }
\!\!
\frac{{{\bar n}\!\!\!/}_{\alpha\beta}}{2}
\langle P| {\bar \chi}_n \, 
\delta_{z_k P,{\cal P}_n}
\delta(\theta^2 - \theta^2_k )
|X 
\rangle_{\alpha} \nn \\ 
&& 
\hspace{5.ex} 
\times 
z_b^{N-1} \frac{{\bar n}\cdot p_k}{P} 
\langle X|\chi_n|P\rangle_\beta  \,.
\eea 
%
 Here $z_b$ is the partonic momentum fraction with respect to the incoming hadron $P$ that enters the hard interaction at the hard scale $Q$. Here $\theta_k$ is the polar angle of the caloremiter $k$ with respect to the beam. 
 The gluon nucleon EEC can be defined similarly using the gauge invariant gluonic field. 
 \begin{figure}[htbp]
  \begin{center}
   \includegraphics[scale=0.2]{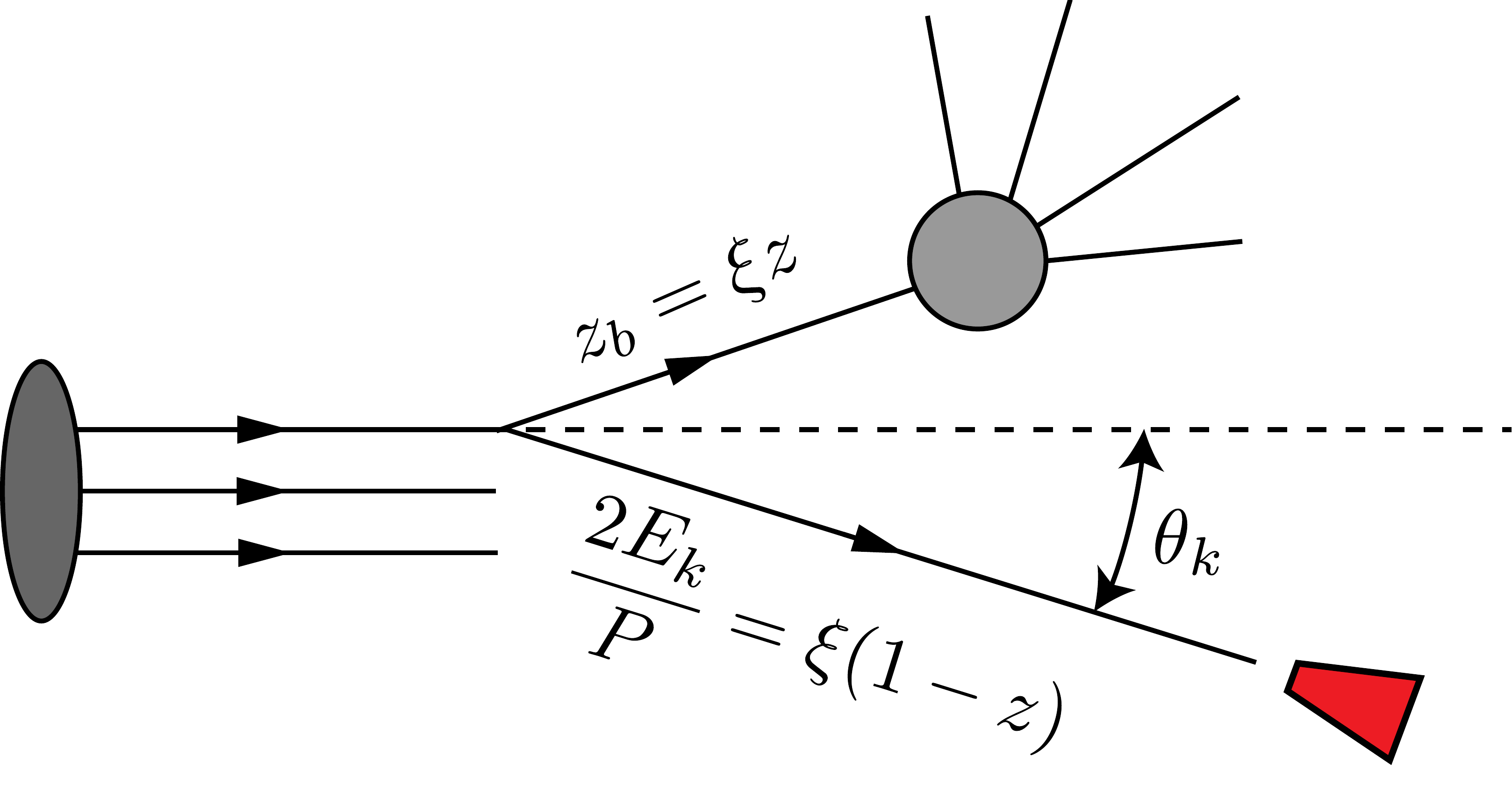} 
\caption{The momentum flow due to an initial state splitting where a fraction of $z_b = \xi z$ goes into the hard interaction at the scale $Q$, while the rest, including the remnants contribute when $\theta_k \sim {\cal O}(\Lambda_{\rm QCD}/Q)$, will deposit in the calorimeter.}
  \vspace{-4.ex}
  \label{fg:split-eec}
 \end{center}
\end{figure}

The nucleon EEC correlates the energy $E_k$ from the initial state radiation that flows into the calorimeter and the energy $z_b$ that participates the hard interaction, as illustrated in Fig.~\ref{fg:split-eec}. 
The nucleon EEC has several fascinating features: 

\noindent 1. The nucleon EEC probes the nucleon internal transverse degrees of freedom through the polar angle $\theta$ when $\theta \sim {\cal O}(\Lambda_{\rm QCD}/Q)$. 
It measures the partonic $\theta$-distribution within the nucleon induced by the intrinsic transverse dynamics.
The nucleon EEC complements the conventional nucleon tomography by using the TMD PDFs. The nucleon EEC evolves as the moment of the collinear PDFs, 
\bea 
\frac{d f_{EEC}^i(N,\theta^2,\mu)}{d\ln\mu^2} =  \gamma^N_{ij}f_{EEC}^j \,, 
\eea 
where $\gamma_{ij}^N$
is the Mellin-moment of the collinear splitting kernel $\gamma_{ij}^N = \int_0^1 dz\, z^{N-1}\, P_{ij}(z) $, and $i\,,j$ are parton flavor indices. 
Compared with the TMD PDFs, the nucleon EEC is free of the Sudakov suppression, 
hence it is likely to provide better resolutions to the intrinsic non-perturbative structures. As concrete examples, we will demonstrate the nucleon EEC can be probed in DIS and will also discuss the generalization of the unpolarized EEC in Eq.~(\ref{eq:beam_eec}) to the polarized case.

\noindent 2.
When the transverse momentum $ \theta Q \gg \Lambda_{\rm QCD}$, the nucleon EEC can be factorized into the product of a perturbatively calculable coefficient $I_{ij}$ and the Mellin moment of the collinear PDF $f_{j/P}(N,\mu) = \int_0^1 d\xi \xi^{N-1} f_{j/P}(\xi,\mu)$, which gives 
\bea\label{eq:factorization} 
f_{EEC}^i (N,\theta^2,\mu) = 
I_{ij}(N,\theta^2,\mu) f_{j/P}(N+1,\mu)\,, 
 \eea 
where $I_{ij}$ encodes the complete information on $\theta$. 
While the details of $I_{ij}$ remained to be calculated order by order using the collinear splitting function similar to the conventional EEC, its scale dependence is fixed to all orders by
\bea\label{eq:dIdlogmu} 
\frac{d I_{ij}(N,\theta^2,\mu)}{d\ln \mu^2 } =  
\gamma_{ik}^{N}I_{kj} - I_{ik} 
\gamma_{kj}^{N+1}  \,.
\eea 
At the leading logarithmic (LL) accuracy, we thus find the $f^i_{EEC}$ satisfies the scaling behavior
%
\bea\label{eq:DL} 
&& f^i_{EEC}(N,\theta^2,\mu)  \nn \\
 = && \left[
e^{\frac{2 \gamma^{(0),N}}{\beta_0}  
L
} I^{(0)}(N,\theta^2)
e^{- \frac{2 \gamma^{(0),N+1} }{\beta_0}   
L
}
\right]_{ij} f_{j/P}(N+1,\mu) \,, 
\eea 
where
$I^{(0)}$ is the leading matching coefficient can be found in the Supplemental Material. 
$\frac{\alpha_2}{2\pi}\gamma_{ij}^{(0),N} $ is the leading order moment of splitting function and $L = \ln \frac{\alpha_s(Q\theta)}{\alpha_s(\mu)}$.
If $\alpha_s(Q\theta)$ is small enough, the nucleon EEC satisfies the scaling behavior $f_{EEC} \sim \theta^{-2}\theta^{ \frac{\alpha_s(Q)}{2\pi} \gamma^{(0),N}} \theta^{  - \frac{\alpha_s(Q)}{2\pi}  \gamma^{(0),N-1} }$. 
In this sense, the nucleon EEC faithfully probes the initial state collinear splitting in the vacuum through the $\theta$ distribution. Deviations from the power-law scaling could shed light on the nature of the initial state source that induces the modification. For example, when then transverse momentum $\theta Q \sim \Lambda_{\rm QCD}$, we expect the non-perturbative structure of proton becomes important, where a non-perturbative modification is needed in \eqref{eq:DL}, whose detailed study we leave for future work. 

\begin{figure}[htbp]
  \begin{center}
   \includegraphics[scale=0.2]{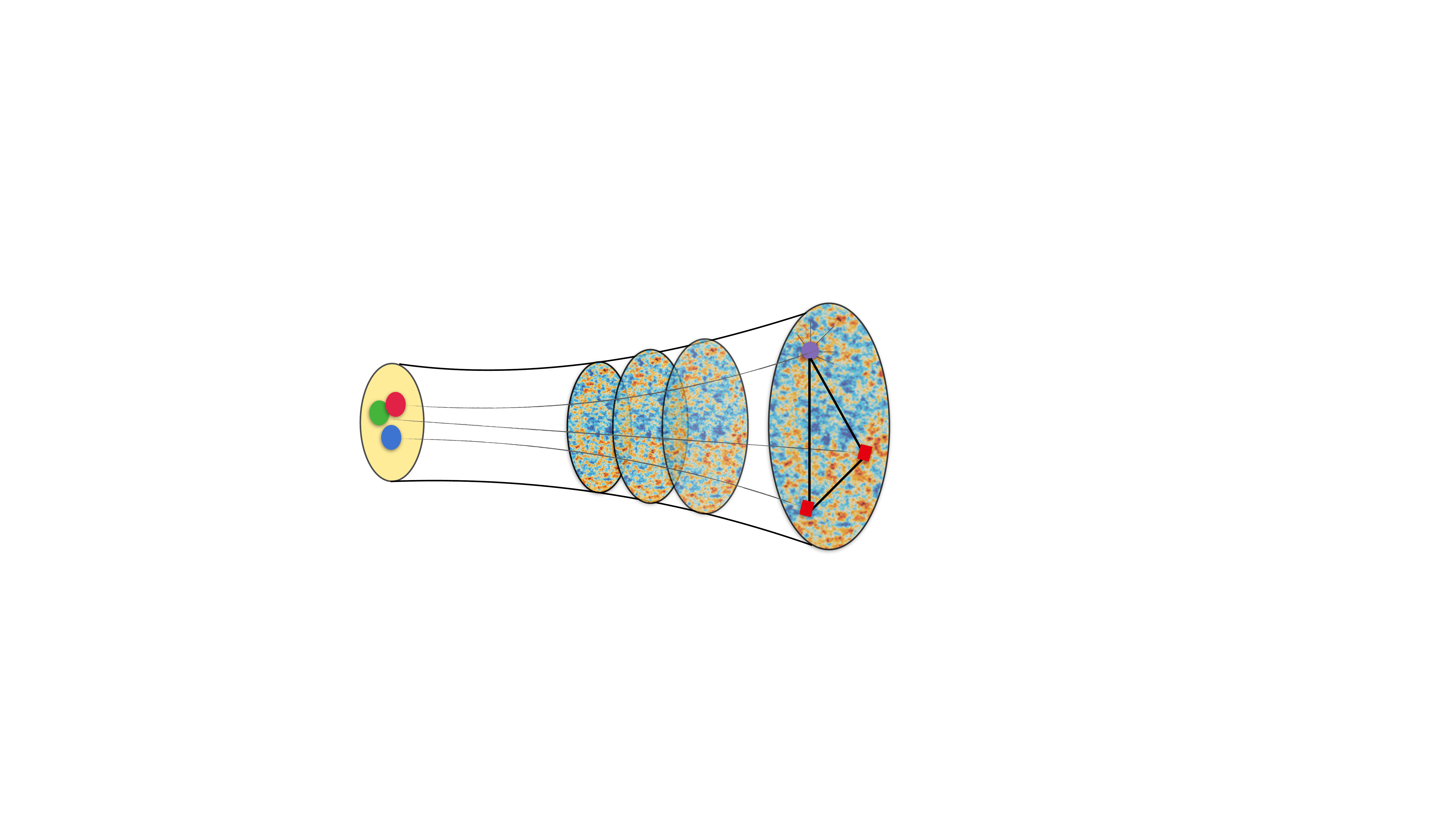} 
\caption{Nucleon 3-point correlation function.}
  \label{fg:3pt-corr}
 \end{center}
   \vspace{-6.5ex}
\end{figure}

\noindent 3. The nucleon EEC can be straightforwardly generalized to the multiple energy correlators by measuring the energy $E_1\,, \dots\,, E_n$ deposit in multiple caloremiters with angular separation $\theta_{ij}$ with $i,j = 1,\dots n$ and the beam.  
The underlying nucleon internal microscopic
details will be imprinted in the detailed structure of these correlation functions, see Fig.~\ref{fg:3pt-corr}, in analogy with the details of the early universe imprinted in cosmological correlation functions. We will leave detailed studies on the nucleon multi-energy correlation to future works. 

\noindent 4. In the extreme small angle limit, we expect simple scaling behavior $\sim \theta^2$ for the nucleon EEC, which signifies the existence of a free hadron phase in initial state beam jet. Similar behavior has been seen in final state jet using CMS open data~\cite{Komiske:2022enw}.

\begin{figure}[!htbp]
  \begin{center}
   \includegraphics[scale=0.25]{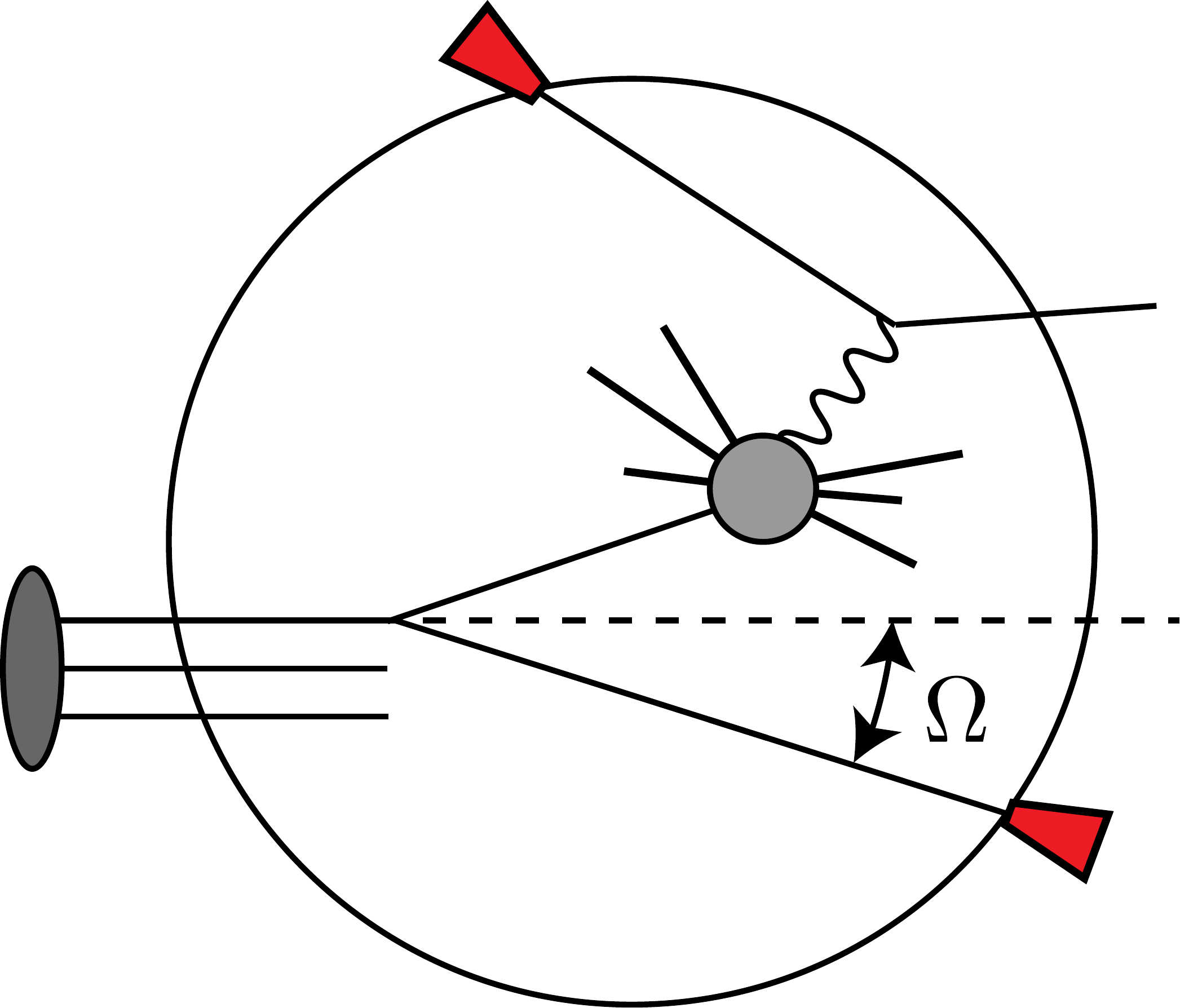} 
\caption{$x_B$ and $p_i$ measurement in DIS that will probe the nucleon EEC $f_{EEC}$. Here $\Omega$ stands for $(\theta_{i},\phi_i)$ with $\phi_i$ the azimuthal angle measured with respect to the nucleon spin.}
  \label{fg:dis-eec}
 \end{center}
   \vspace{-4.ex}
\end{figure}

\textbf{\emph{The $x_B$ weighted deep-inelastic scattering.}} To see how the nucleon EEC can be measured, we consider the DIS process $l + P \to l' + X $ in the frame where the virtual photon $\gamma^\ast$ acquires no transverse momentum, e.g., the $\gamma^\ast$-$P$ center of mass frame or the Breit frame. We assume the nucleon is moving along the $+z$-direction. 
We measure the Bjorken $x_B = \frac{-q^2}{2P\cdot q }$ and the momentum flow $p_i^\mu$ deposit (including the ones from the beam remnants) in a calorimeter along the direction $i$. Here $q = l'-l$ is the momentum carried by the virtual photon. In this manuscript, we will be particularly interested in the scenario where the detector is placed in the far-forward region and therefore the transverse momentum flow $p_{i,t}^2 \sim \theta^2 E_i^2$ is very small compared with $q^2 = -Q^2$, as depicted in Fig~\ref{fg:dis-eec}. 

The measurement probes the weighted cross section $\Sigma_N(Q^2,\theta)$ in Eq.~(\ref{eq:eec-def1}).
%
%
%
%
We note that the polar angle $\theta_i$ is 
related to the transverse momentum as
$\sin \theta_i = \frac{p_{i,t}}{E_i}$. For the time being, we assume the proton is unpolarized.


The weighted cross section in Eq.~(\ref{eq:eec-def1}) can be calculated via 
\bea\label{eq:eec-def2}
&& \Sigma_N(Q^2,\theta^2)
=  \frac{\alpha^2 }{Q^4}
\int dx_B x_B^{N-1}
\!\!\!
\sum_{{i\in X},{\lambda=T,L} }
\!\! e_i^2
f_{\lambda}(y) 
\, 
\nn \\ 
&&   \times
\int 
d^4x
e^{i q \cdot x }
\,  
\langle P |j^\dagger\!\! \cdot\! \epsilon^\ast_\lambda   \, 
\frac{  2 {\cal E}(\theta)}{P} \, 
j \! \cdot \! \epsilon_\lambda(x) | P \rangle 
\,, \quad 
\eea 
where $j^\mu$ is the conserved current. $\epsilon_\lambda$ is the virtual photon polarization vector with $\lambda = L,T$ for the longitudinal and transverse polarization, respectively. $f_\lambda(y)$ is the photon flux such that $f_T = 1-y+y^2/2$ and $f_L = 2(1-y)$ where $y = \frac{2p\cdot q}{2p\cdot l}$ is the inelastcity. We note that the property of the similar matrix element with $|P\rangle$ replaced by the vacuum state has been discussed in context of the conformal collider physics~\cite{Hofman:2008ar}. 

When $\theta \ll 1$ and thus $i$ is close to the beam, it is ready to show by using SCET~\cite{Bauer:2000yr,Bauer:2001yt,Bauer:2001ct,Beneke:2002ph,Bauer:2002nz} 
 that $\Sigma_N$ takes the factorized form at LL
\bea\label{eq:eec-fact} 
\Sigma_N(Q^2,\theta^2) = 
f^{i}_{EEC}(N,\theta^2,\mu) \, 
\int d \zeta  \zeta^{N-1}   \, 
\frac{d^2 \hat{\sigma}_i(\mu)}{d\zeta dQ^2}    
+ {\cal O}(\theta) 
\,, \quad
\eea 
where we see the occurrence of the nucleon EEC $f^i_{EEC}(N,\theta)$ and therefore the proposed measurement does probe the nucleon EEC. The $z_b^{N-1}$ within $f^i_{EEC}(N,\theta)$ in   Eq.~(\ref{eq:beam_eec}) enters through the $x_B^{N-1}$ weight. 
We note that the coefficient of $f_{EEC}^i$ is nothing but the Mellin-moment $d\hat{\sigma}_i(N,\mu)$ of the partonic DIS cross section~\footnote{Strictly speaking, this is only true if we integrate $x_B$ down to $0$ in Eq.~(\ref{eq:eec-def1}). However, if we choose sufficiently large $N$, to suppress the contribution from the small values of $x_B$, the integral will be well approximated by the Mellin-moment.}, satisfy $d\hat{\sigma}_i(N,\mu)/d\ln\mu^2
= - \gamma^N_{ij} \hat{\sigma}_j(N,\mu)
$. Here $i$ and $j$ can either be a quark or gluon. 

When $\theta Q \gg \Lambda_{\rm QCD}$, the nucleon EEC is further factorized following Eq.~(\ref{eq:factorization}). Thus the scale dependence of the coefficient $I_{ij}$ in Eq.~(\ref{eq:dIdlogmu}) is an immediate consequence of the scale independence of the weighted cross section $d\Sigma_N/d\ln \mu = 0$. 

Now we estimate the requirement of the forward detector for this measurement. Suppose we want to probe the intrinsic transverse momentum of the nucleon, we will demand the detector to detect transverse momentum flow $p_{i,t} \sim \Lambda_{\rm QCD}$ and thus to cover polar 
angles down to $\theta \sim p_{i,t}/ Q$. Hence for $Q\sim {\cal O}(5\>{\rm GeV})$, the estimated $\theta \sim {\cal O}(0.2\> \text{rad})$, which is well covered by the EIC far-forward particle detection plan~\cite{AbdulKhalek:2021gbh,Proceedings:2020eah,Arratia:2022quz,Cebra:2022avc} and will be even better favored if the coverage proposals such as the Zero Degree Calorimeter~\cite{Arratia:2022quz,Bylinkin:2022rxd} down to and below $5\,{\rm mrad}$ would be realized. We emphasize that since we only count the energy deposit in the calorimeters, no jet clustering procedure is needed. Meanwhile, instead of using the calorimetry, the track-based measurements can be carried out~\cite{Chen:2020vvp,Li:2021zcf,Jaarsma:2022kdd} to offer better pointing and angular resolution. 

Here we predict the normalized differential distribution $\langle \text{EEC} \rangle_N = 1/\sigma \theta^2 d \Sigma_N(\theta^2)  $ in the Breit frame. We define the rapidity $y = \ln \tan \theta/2$. For the prediction, we use {\tt Pythia82}~\cite{Sjostrand:2014zea} with the proton $P = 275\>{\rm GeV}$ and the incoming lepton $l = 10\>{\rm GeV}$. 
 \begin{figure}[htbp]
  \begin{center}
   \includegraphics[scale=.7]{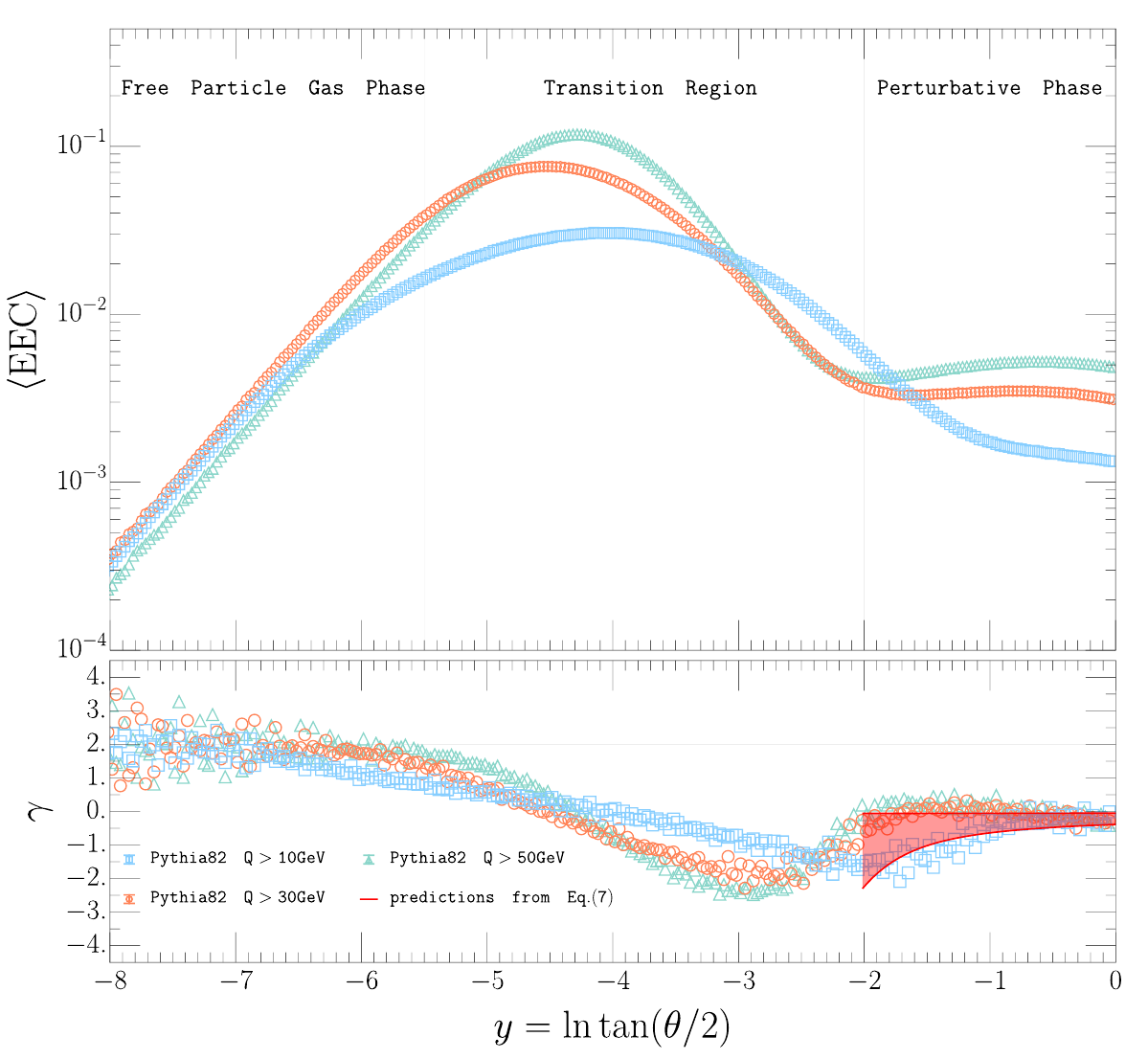} 
\caption{$\langle \text{EEC} \rangle$ and $\gamma$ distribution in the Breit Frame.}
  \label{fg:eic-eec}
 \end{center}
   \vspace{-5.ex}
\end{figure}

In Fig.~\ref{fg:eic-eec}, we show the predictions for $\langle \text{EEC} \rangle \equiv \langle \text{EEC} \rangle_2$, i.e., with $N=2$. 
We vary the values of $Q$ with $Q>10\>{\rm GeV}$, $Q>30\>{\rm GeV}$, $Q>50\>{\rm GeV}$. We see from the upper panel of Fig.~\ref{fg:eic-eec} that although the $Q$'s are different, the predicted $\langle \text{EEC} \rangle $'s display similar features, which implies that the normalized distributions reflect the property of the nucleon itself at different scales $\mu \sim Q$. 

We note that Fig.~\ref{fg:eic-eec} exhibits an interesting ``phase transition" between the perturbative-phase for $\theta \gtrsim 0.2\> {\rm rad}$ and the ``free-particle-phase" for $\theta \lesssim 0.005\> {\rm rad}$, connected by the non-perturbative transition region. In the perturbative region, the distribution is almost flat, largely independent of $Q$'s, which is a direct manifestation of Bjorken scaling in the space of polar angle~(rapidity), as can be seen from Eq.~\eqref{eq:DL}.

The feature is more evident by looking at the slope $\gamma$ showed in the lower panel of Fig.~\ref{fg:eic-eec}, where in the perturbative region $\gamma \sim 0$, while in the deep non-perturbative region for $\theta \lesssim 0.005\> {\rm rad}$, $\gamma \sim 2$. It will be very interesting if we can confirm such phase transition at future experimental facilities. 

We again notice that all the slope $\gamma$ distributions with different values of $Q$ shares similar behaviour which indicates it reveals the intrinsic property of the nucleon EEC at different scales $\mu$. The transition region moves to the right as we decrease $Q$, which is expected since the transition occurs when $\theta \sim {\cal O}(\Lambda_{\rm QCD}/Q)$. The $\gamma$ in the perturbative region can be predicted using the LL result in Eq.~(\ref{eq:DL}) and the factorization in Eq.~(\ref{eq:eec-fact}), which is showed in the red line. All the $Q$ values are covered within the band, obtained by a dramatic variation of $\mu$, from $\mu = 50\> {\rm GeV}$ to $\mu = 300\, {\rm MeV}$. We find good agreement between the {\tt Pythia} simulations and the analytic LL result in the perturbative phase. We emphasize that 
future observed deviation from the predicted slope could be used to extract the nature
of the initial state source that induces the modification to the collinear splitting kernel, such as the hadronization and the hot/cold nucleus medium effects. The theory precision can be further improved and we leave it to future works.

\begin{figure}[htbp]
  \begin{center}
   \includegraphics[scale=.28]{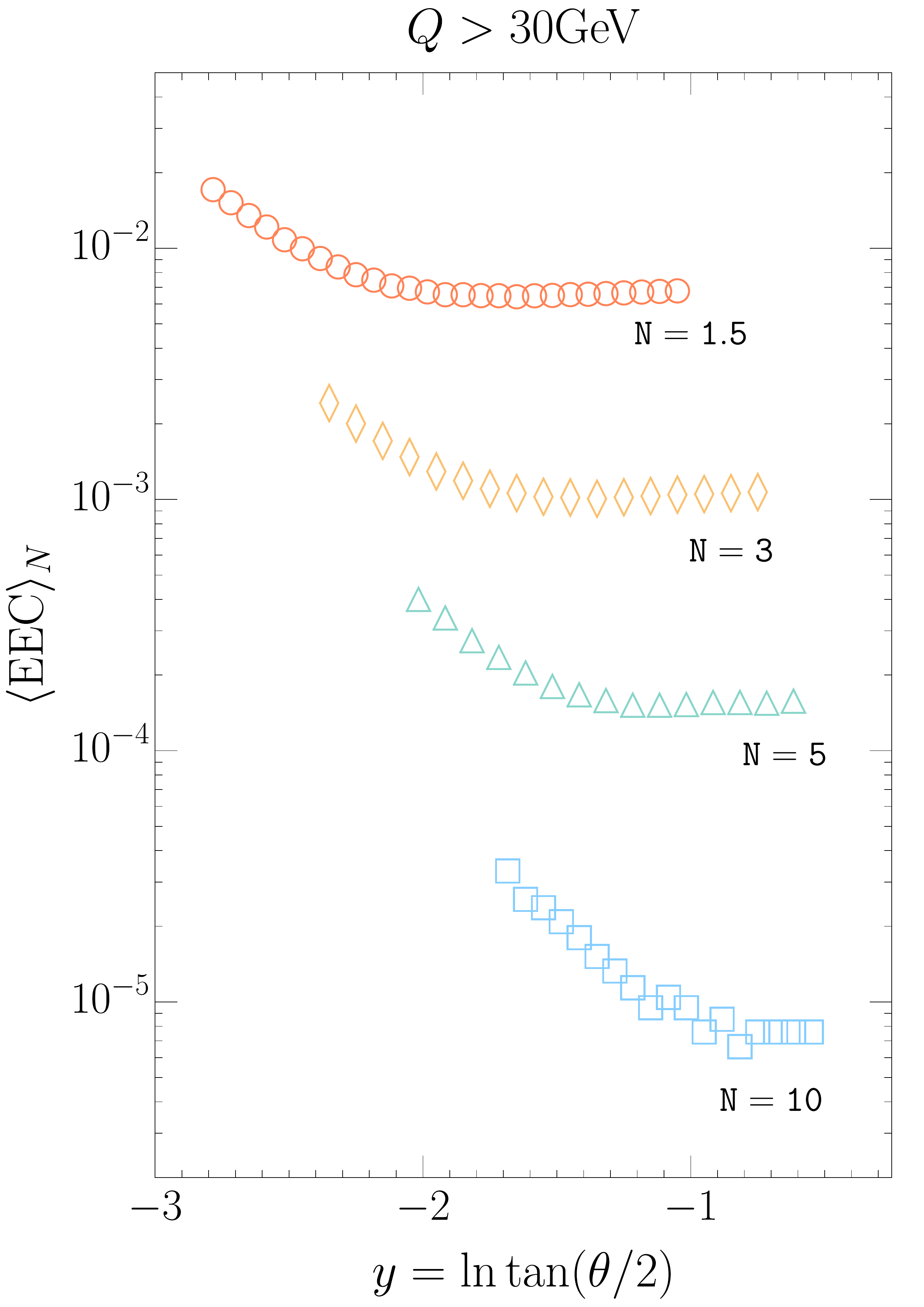} 
\caption{ The angular ``Bjorken scaling rule" of the nucleon EEC. }
  \label{fg:angular-bjorken-scaling}
 \end{center}
   \vspace{-5.ex}
\end{figure}

In Fig.~\ref{fg:angular-bjorken-scaling}, we show the $\langle \text{EEC} \rangle_N$ for different $N$. The smaller values of the $N$ increase the sensitivity to the small-$x$ component of a nucleon while larger $N$ probes more into the large $x$ regime. 
Again the plot manifests the polar angle (rapidity) scaling behavior of the nucleon EEC, similar to the famous Bjorken scaling rule of the PDFs. We note that for larger $N$, therefore effectively larger $x$ closer to $1$, the transition region appears at larger angle, consistent with the expectation that $\theta \sim {\cal O}(\Lambda_{\rm QCD}/Q)$ increase as $Q$ decrease. Also since the normalization of nucleon EEC is propotional to moment of PDFs, we expect to see a decrease in the magnitude as $N$ increase, as a result of small-$x$ enhancement of PDFs.



\textbf{\emph{Transversely polarized EEC.}} Once the incoming nucleon is polarized, we can also probe the spin asymmetry by measuring the azimuthal modulation. Here the beam and the out-going lepton span the $x$-$z$ plane.  

For instance, for the transversely polarized nucleon beam, if we measure the $\sin(\phi-\phi_S)$ distribution with $\phi-\phi_S$ the azimuthal angle between the detector and the nucleon spin, we are probing the spin dependent distribution $d \Sigma_N(\vec{n}_t,S_T) $ whose $\sin(\phi-\phi_S)$  dependent part factorized similarly as the unpolarized case in Eq.~(\ref{eq:eec-fact}), with the replacement of $f^i_{EEC}(\theta)$ by  
\bea\label{eq:beam_eec_pol}
&& \frac{-\epsilon_T^{a b}n^a_{t}S_T^b}{M_P}\, 
f^q_{T,EEC}(N,\theta ) = 
\sum_X \sum_{i \in X} z_b^{N-1} \frac{{\bar n}\cdot p_i }{P} 
\frac{{{\bar n}\!\!\!/}_{\alpha\beta}}{2}
  \nn \\
&& \times  
\langle P,S_T| {\bar \chi}_n 
\delta_{z_i P,{\cal P}_n}
\delta^{(2)}(\vec{\hat{n}}_{t} - \vec{\hat{n}}_{i,t} ) 
|X 
\rangle_{\alpha}  \langle X|\chi_n|P,S_T\rangle_\beta
 \,, \qquad 
\eea 
for quark, 
where $\vec{n}_t = \sin\theta( \cos\phi,   \sin\phi )$, $S_T$ is the nucleon spin and $M_P$ is the nucleon mass. The non-vanishing of $f_{T,EEC}$ is owing to the same mechanism that gives rise to the Sivers effect~\cite{Sivers:1989cc,Collins:2002kn}. The Sivers-like EEC $f_{T,EEC}$ induces the $\sin(\phi-\phi_S)$ azimuthal asymmetry 
$A_N = \frac{d\Sigma_N(S_T) - d\Sigma_N(-S_T)}{\Sigma(S_T)+\Sigma(-S_T)}$.
The predicion of $A_N$ relies on the non-perturbative input of the $f_{T,EEC}$ which requires further studies in the future. Since there is no Sudakov surprresion, we anticipate a better chance to observe the asymmetry at the EIC.  

\textbf{\emph{Conclusion.}} In this manuscript, we introduced the nucleon energy-energy-correlator (EEC) that measures the  correlation of the energy flows from the initial nucleon. The nucleon EEC reflects the parton angular distribution in a nucleon. 
This new object is novel both theoretically and phenomenologically. Theoretically, 
we have demonstrated that the microscopic details of the nucleon such as the faithful vacuum collinear splitting behavior as well as the nucleon internal transverse momentum and spin degrees of freedom are imprinted in the energy correlation function, and meanwhile the nucleon EEC may offer additional possibilities to understand the nucleon structures using the light-ray OPE in QCD. Phenomenologically, we showed how the nucleon EEC can be probed at EIC with a far-forward detector. We have set the theoretical foundation for the observable and predict the measured distribution at EIC to exhibit power law scaling behaviours. A novel phase transition between the free-particle and the perturbative phases is observed. In the perturbative region, the polar angle (rapidity) version of the Bjorken scaling behavior is also predicted.  

One advantage of the nucleon EEC is that its measurement involves no jet clustering procedure nor additional non-perturbative object other than the nucleon EEC itself. 
Besides, the factorization in Eq.~(\ref{eq:eec-fact}) only involves a product instead of a convolution should make the extraction of the nucleon EEC a lot easier. Therefore it serves a clean complement to the conventional TMDs to the nucleon structures, good for either high energy or low energy machines. 

Other than the scenarios considered in the manuscript, we expect the proposed nucleon EEC to have a wide application to future nucleon/nucleus studies. Extensions to other observables sensitive to the various TMD distributions will follow straightforwardly. By suitably choosing the weight $N$, the nucleon EEC can be made sensitive to the small-$x$ phenomenology. The nucleon EEC can also be used to study the cold nuclear effect in $eA$ collisions or to extract the hot medium effect with heavy ion data. All these effects will leave its foot print in the deviations from the $\langle \text{EEC} \rangle_N $ and its slope introduced in this work. 
As long as one charm is tagged in the detected forward event, the nucleon EEC can offer a direct look into the intrinsic charm content.  
Furthermore, the generalization of the EEC to multiple point correlations will allow for more delicate  differentiation of the nucleon/nucleus microscopic details. 
We thus anticipate that
the nucleon EEC introduced in this work will stimulate further
theoretical developments along these directions. 

\begin{acknowledgments}
{\it Acknowledgement.} We are grateful to Miguel Arratia, Hao Chen, Zhong-bo Kang, Ian Moult, Jinlong Zhang, Jian Zhou for insightful discussions. We thank for the hospitality of the committee for the ``Heavy flavor and QCD" workshop held in Changsha where this work was initiated. We thank for stimulating feed-backs from the EicC bi-week meeting. This work is supported by the Natural Science Foundation of China under contract No.~12175016 (X.~L.), No.~11975200 (H.~X.~Z.).
\end{acknowledgments}

\appendix
\begin{widetext}

\section{Supplemental Materials for ``The Nucleon Energy Correlators"}

In this supplemental material, we present the operator definition of the nucleon energy-energy correlator (EEC) while the factorization theorem that gives rise to the nucleon EEC will be given else where~\cite{future:2023tap}. We comment the feature of the nucleon EEC, demonstrate the absence of the perturbative Sudakov factor and highlight the calculation of the quark energy-energy correlator $f_{\rm EEC}$ at ${\cal O}(\alpha_s)$ for $\theta Q > \Lambda_{\rm QCD}$. The detailed calculations of all the $f_{\rm EEC}$'s, including their higher order generalization,  will be presented in the on-going work~\cite{future:2023tap}.
We also present a model calculation to illustrate the relation of $f_{\rm EEC}$ to the Trnasverse-Momentum-Dependent~(TMD) PDFs.

\subsection{operator definition and one-loop calculation}

The operator definition of the nucleon energy-energy correlator (EEC)
with both $\theta$ and $\phi$ dependence is given by
\bea\label{eq:ope-def} 
&& f_{q,{\rm EEC}}(N,\theta,\phi)
=
\int_0^1 z^{N-1} \int \frac{dy^-}{2\pi } e^{- i z P^+ y^- }
\left\langle P \left|
{\bar \chi}_n(y^-) \frac{\gamma^+}{2} 
\frac{2{\cal E}(\theta,\phi)}{P^+}   \chi_n(0)
\right| P \right\rangle  \,, \nn \\ 
&&f_{g,{\rm EEC}}(N,\theta,\phi) 
= - 
\int_0^1 z^{N-1} 
\int \frac{dy^-}{2\pi z P^+} e^{- i z P^+ y^- }
\left\langle P \left| {\bar n}\cdot {\cal G}_\perp (y^-)  
\frac{2{\cal E}(\theta,\phi)}{P^+}
{\bar n}\cdot {\cal G}_\perp (0)  \right|P \right\rangle  \,, 
\eea 
where $\chi$ and ${\cal G}_\perp$ are the SCET gauge invariant collinear quark and gluon fields, respectively~\cite{Bauer:2000yr,Bauer:2001yt,Bauer:2001ct,Beneke:2002ph,Bauer:2002nz}. The transverse angular information of the nucleon enters through the energy operator ${\cal E}(\theta,\phi)$ that measures the energy deposit in the detector at given angles $\theta$ and $\phi$~\cite{Sveshnikov:1995vi,Tkachov:1995kk,Korchemsky:1999kt,Bauer:2008dt}, 
\bea 
{\cal E}(\theta,\phi) |X\rangle = 
\sum_{i\in X} E_i \delta(\theta_i^2 - \theta^2 )  \delta(\phi_i - \phi ) | X\rangle \,.  \eea 

Here are some comments on the nuclone EEC $f_{\rm EEC}$: 
\begin{itemize} 
\item We note that the soft contribution with momentum scaling $p_s =  (p_s^+,p_s^-,p_{s,t}) \sim Q(\theta,\theta,\theta)$  to the EEC will be power suppressed and absent from $f_{\rm EEC}$, since ${\cal E}(\theta,\phi) |X_s\rangle = E_s\delta(\theta^2_s-\theta^2)\delta(\phi_s-\phi) |X_s \rangle \sim \theta Q |X_s \rangle \to 0$ as $\theta \to 0$, compared with the collinear contribution with $p_c \sim Q(1,\theta^2,\theta)$ and thus  
${\cal E}(\theta,\phi) |X_c\rangle = E_c\delta(\theta^2_c-\theta^2)\delta(\phi_c-\phi) |X_c \rangle \sim Q|X_c\rangle $. The absence of the soft contribution indicates that there will be no perturbative Sudakov logarithms in $f_{\rm EEC}$ and thus no perturbative Sudakov suppression factor. 

 \begin{figure}[htbp]
  \begin{center}
   \includegraphics[scale=0.35]{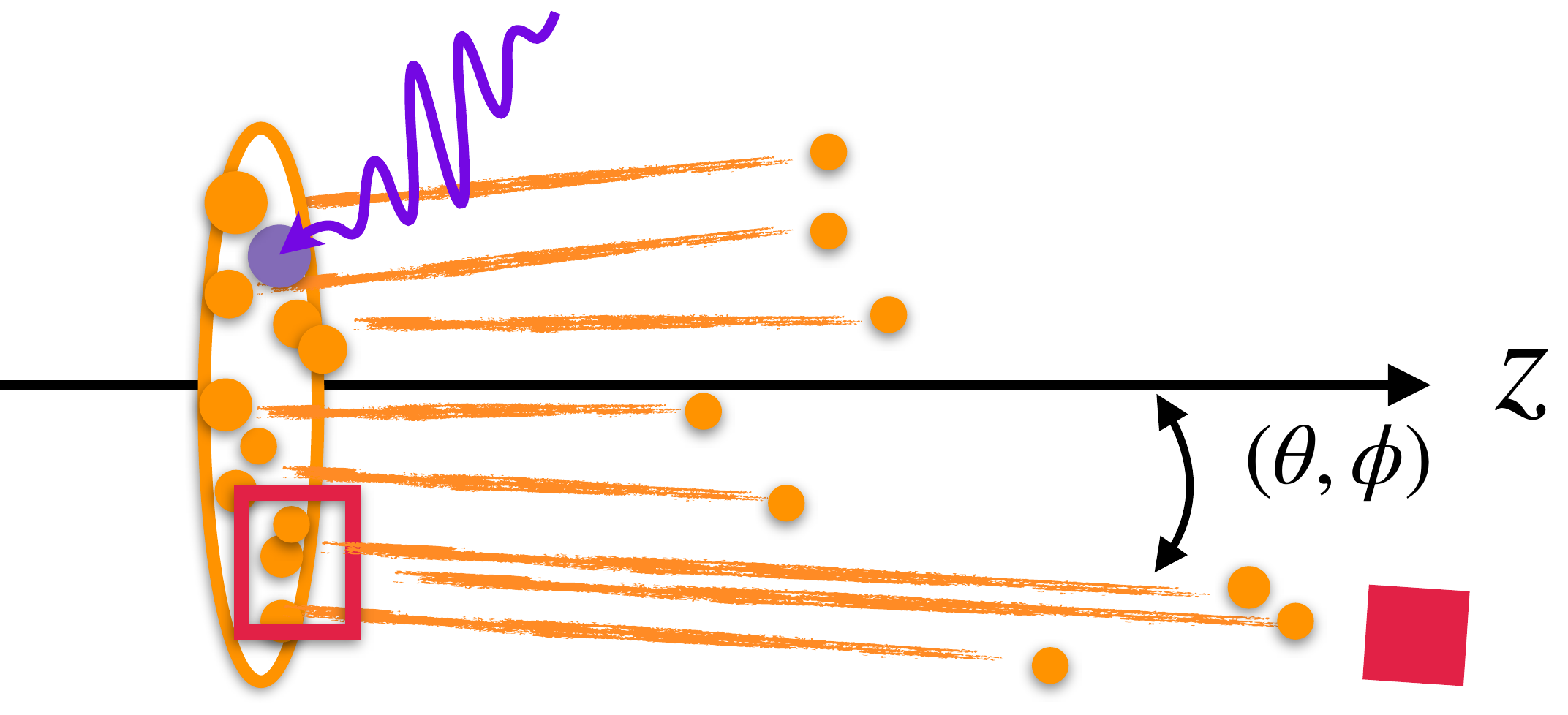} 
\caption{When $Q\theta \sim \Lambda_{\rm QCD}$, the partons in the proton with intrinsic transverse momentum $\frac{p_i^+}{2} \theta (\cos \phi, \sin \phi)$ are detected through the ${\cal E}(\theta,\phi)$. By looking at the detectors (represented by the red block) at different $(\theta,\phi)$, we are effectively imaging the internal structure of the proton in the azimuthal plane where $\theta$ dertmines the angular distance to the $z$-axis.}
  \vspace{-4.ex}
  \label{fg:eec-image}
 \end{center}
\end{figure}
%

\item The internal structure of the proton or more specifically the partonic radiation angular $(\theta,\phi)$ distribution enters through ${\cal E}(\theta,\phi)$, when $\theta Q \sim \Lambda_{\rm QCD}$, as illustrated in Fig.~\ref{fg:eec-image}, where the intrinsic transverse dynamics of the partons generate the non-zero angles $(\theta,\phi)$ and drive the partons to hit the detector. In this way, by recording the energy deposit in the detectors (realized by the energy flow operator ${\cal E}(\theta,\phi)$) located at different $(\theta,\phi)$, we are able to 
scan the internal structure of the proton.

\item The definition can be generalized to multiple correlators 
\bea\label{eq:multi-def} 
&& f_{q,{\text{E$^n$C}}}(N,\{\theta_i,\phi_i\})
=
\int_0^1 z^{N-1} \int \frac{dy^-}{2\pi } e^{- i z P^+ y^- }
\left\langle P \left|
{\bar \chi}_n(y^-) \frac{\gamma^+}{2} 
\frac{2{\cal E}(\theta_1,\phi_1)}{P^+} \dots
\frac{2{\cal E}(\theta_n,\phi_n)}{P^+}
\chi_n(0)
\right| P \right\rangle  \,, \nn \\ 
&&f_{g,{\text{E$^n$C}}}(N,\{\theta_i,\phi_i\})
=
\int_0^1 z^{N-1} 
\int \frac{dy^-}{2\pi z P^+} e^{- i z P^+ y^- }
\left\langle P \left| {\bar n}\cdot {\cal G}_\perp (y^-)  
\frac{2{\cal E}(\theta_1,\phi_1)}{P^+}\dots 
\frac{2{\cal E}(\theta_n,\phi_n)}{P^+}
{\bar n}\cdot {\cal G}_\perp (0)  \right|P \right\rangle  \,,  \qquad
\eea 
which measures the correlation among the partons of the proton sit at different angular positions in the azimuthal plane. The multiple correlators will help to establish a more detailed differential picture of the proton structures.

\item When $\theta Q \gg \Lambda_{\rm QCD}$, the $f_{\rm EEC}$ can be calculated perturbatively. 
Here, we show one example that contributes to the $\Sigma_N$ at order $\alpha_s$, to highlight in detail how such calculation is performed. 
We  considered the case shown in Fig.~\ref{fg:eec-feyn}, in which an incoming gluon $g$ from the proton splits into a $q{\bar q}$ pair, one of which enters the hard interaction and the other hits the detector at $\theta$. Other channels can be obtained similarly~\cite{future:2023tap}.   
\begin{figure}[htbp]
  \begin{center}
   \includegraphics[scale=0.85]{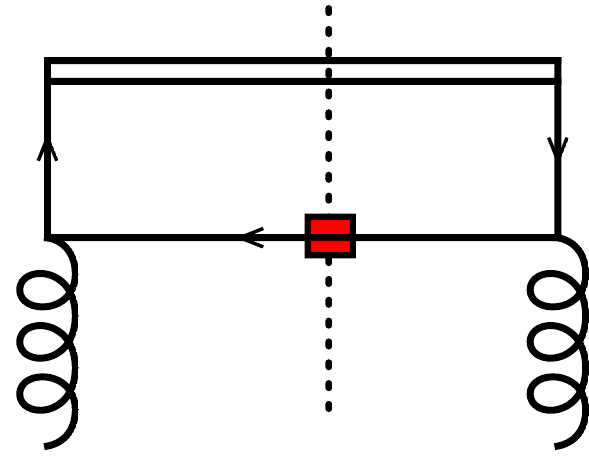} 
\caption{Representative contribution to the $f_{q,{\rm EEC}}$ initiated by a gluon with momentum $\xi P$ out of the proton splitting into $q{\bar q}$ with momentum $p$ and $q$, respectively. The vertical dashed line stands for the phase space cut, the red block for the ${\cal E}(\theta)$ insertion and the double line is the Wilson line.}
  \vspace{-4.ex}
  \label{fg:eec-feyn}
 \end{center}
\end{figure}

Fig.~\ref{fg:eec-feyn} contributes to the quark nucleon EEC $f_{q,{\rm EEC}}$ as
\bea 
f_{q,{\rm EEC}}(N,\theta) = \int dz z^{N-1}  \sum_X \mu^{2\epsilon}
\int \frac{d^d q}{(2\pi)^{d-1}} \delta(q^2) 
\delta((1-z) P^+-q^+ - p_X^+) 
\frac{q^+}{P^+} 
\delta(\theta^2_q - \theta^2)
 \langle P | {\bar \chi}(0)  |{\bar q} X \rangle \langle {\bar q} X | \frac{\gamma^+}{2} \chi |P \rangle   \,, \quad 
\eea 
where we have inserted the complete set $\int \frac{d^d q}{(2\pi)^{d-1}} \delta(q^2)  \sum_X |{\bar q}X\rangle \langle {\bar q} X |$ into the operator definition in Eq.~(\ref{eq:ope-def}) and applied the translation operation on $y^-$. At the order $\alpha_s$, the final state ${\bar q}$ is from the gluon splitting, while $X$ stands for the remnants from the proton. For $\theta  > 0$, we safely set $\epsilon=0$ and thus $d = 4$. The gluon splitting can be calculated using the SCET Feynman rules, see for instance, Fig.~1 in~\cite{Bauer:2000yr}, which gives   
\bea\label{eq:split}
f_{q,{\rm EEC}}(N,\theta) &=& 
 \int dz z^{N-1}
\frac{1}{2}
\int \frac{dq^+ d^{2}q_t}{q^+(2\pi)^3}  
\frac{q^+}{P^+} 
\delta((1-z) P^+-q^+ - p_X^+) 
\delta(\theta^2_q - \theta^2)\nn\\
&& \times g_s^2 {\rm Tr}[T_aT_a]
\frac{1}{2}
{\rm Tr}[\gamma^\mu_t \gamma_{t,\mu} {q {\!\!\!  /}} \gamma^+ ] \left( 
\frac{ {-\vec{q}^2_t} }{(q^+)^2} 
+ \frac{ {-\vec{p}^2_t} }{(p^+)^2}  
\right)
\left(
\frac{p^+}{p^2} \right)^2 \nn \\ 
&& \times \frac{1}{2}\frac{1}{N_C^2-1}\sum_X
\langle P|
\, A_{t,a,\mu}|X\rangle \langle X| A_t^{a,\mu} |P \rangle  \int d\xi P^+ \delta((1-\xi)P^+-p_X^+) \,, 
\eea 
where the subscript ``$t$" stands for the transverse component. We have inserted in the last line $1 = \int d\xi P^+ \delta((1-\xi)P^+ - p_X^+)$ to define the variable $\xi$, and we averaged over the initial color and polarizations.

To proceed, we note that since the intrinsic transverse momentum of the gluon, $l_t \sim \Lambda_{\rm QCD}$ and is negligible when $\theta Q \gg \Lambda_{\rm QCD}$, we have $\vec{p}_t = - \vec{q}_t$ and 
\bea 
p^2 = (l-q)^2 = - 2l\cdot q = -  \frac{l^+}{q^+} \vec{q}_t^2 \,. 
\eea 
Working out the traces, we find
\bea 
f_{q,{\rm EEC}}(N,\theta) &=& 
 \int_0^1 dz z^{N-1}  \int_z^1 \frac{d\xi}{\xi } 
\, 
\frac{1}{2} \frac{ 1 }{8\pi^3}
\int \frac{dq^+ d^{2}q_t}{\xi P \, q^+  }  
\frac{q^+}{P^+} 
\delta\left(1- \frac{z}{\xi} - \frac{q^+}{\xi P^+} \right) 
\delta(\theta^2_q - \theta^2)\nn\\
&& \times 4\pi\alpha_s T_R
2 q^+ \frac{1}{\vec{q}_t^2}
\left( 
\frac{ (p^+)^2 }{(q^+)^2} 
+ 1   
\right)
\left(
\frac{q^+ }{g^+} \right)^2 \nn \\
&& \times \sum_X
(-)
\langle P|
\, A_{t,a,\mu}|X\rangle \langle X| A_t^{a,\mu} |P \rangle  (\xi P^+) \delta((1-\xi)P^+-p_X^+) \,.
\eea 
where we notice that the last line gives nothing but the gluon PDF, see for instance~\cite{Stewart:2010qs}. Working out the $q^+$ and the solid angle integration, we find 
\bea 
f_{q,{\rm EEC}}(N,\theta) &=& 
 \int \frac{d\xi}{\xi } \int dz z^{N-1}  
\left(1 - \frac{z}{\xi} \right) \xi 
\int \frac{ dq^2_t}{  q_t^{2}    } 
\delta(\theta^2_q - \theta^2)  \nn \\ 
&& \times \frac{\alpha_s}{2\pi} T_R  
\left(
\left( \frac{z}{\xi} \right)^2
+ \left(1- \frac{z}{\xi}  
\right)^2
\right)
f_{g/P}(\xi) \,.
\eea 
Now we make the varaible change $\frac{z}{\xi} \to x$ and use the fact that $q_t = 
q_z \sin \theta =
\frac{q^+}{2}\theta_q$ for small $\theta_q$, to find
\bea 
\label{eq:coll}
f_{q,{\rm EEC}}(N,\theta) &=& 
 \int dx\, x^{N-1} (1-x) 
\,  \frac{\alpha_s}{2\pi} T_R   \Bigg\{ 
    \frac{1}{\theta^2} 
\left[  x^2+(1-x)^2 \right] 
  \Bigg\}
\int d\xi \xi^{N}
f_{g/P}(\xi) \,
\eea 
We notice that the result factorized into product of a matching coefficient and the moment of PDF. The angular dependence $1/\theta^2$ is governed by the splitting function $P_{qg}(x)$.
The calculation also suggests that $ \frac{q^+}{2}\theta$ probes the partonic transverse momentum $q_t$. The $q\to q$ channel can be calculated in the same way. 

Once sum over all partonic channels, we find the quark nucleon EEC  
\bea 
f_{q,{\rm EEC}}(N,\theta^2) &=& 
I^{(0)}_{qg}(N,\theta^2) f_{g/P}(N+1) + I^{(0)}_{qq}(N, \theta^2) f_{q/P} (N+1) \,, 
\eea 
with 
the coefficient $I^{(0)}_{qg}$ given by 
\bea 
I^{(0)}_{qg}(N,\theta^2) &=& 
 \int dx\, x^{N-1} (1-x) 
\,  \frac{\alpha_s}{2\pi} T_R     \frac{1}{\theta^2}  
\left[  x^2+(1-x)^2 \right]
= 
  \frac{\alpha_s}{2\pi}     
\Big(\gamma_{qg}^{(0)} (N) - \gamma_{qg}^{(0)} (N+1) \Big) \frac{1}{\theta^2}
\,,
\eea 
where 
\begin{equation}
    \gamma_{qg}^{(0)}(N) = T_R  \frac{2 + N + N^2}{N (N+1) (N+2)} \,. 
\end{equation}

Following the same line, the leading matching coefficient for $I^{(0)}_{qq}(N, \theta^2)$ is given by
\bea\label{eq:Iqq}
I^{(0)}_{qq}(N,\theta^2) &=& 
 \int dx\, x^{N-1} (1-x) 
\,  \frac{\alpha_s}{2\pi} C_F  
\frac{1}{\theta^2}\left[ \frac{ 1 + x^2}{1 - x}\right]
\,.
\eea
We can see explicitly from Eq.~(\ref{eq:Iqq}) that the $(1-x)$ factor originated from the energy weighting will cancel the soft divergent term $1/(1-x)$ when $x\to 1$. Therefore there is no perturbative Sudakov logarithm in $f_{\rm EEC}$ as we have argued previously. 

Perform the $z$ integral we find 
\bea 
I^{(0)}_{qq}(N,\theta^2) &=& 
\,  \frac{\alpha_s}{2\pi} \left( \gamma_{qq}^{(0)}(N) -  \gamma_{qq}^{(0)}(N+1) \right) \frac{1}{\theta^2}  \,, 
\eea    
where 
\bea 
\gamma_{qq}^{(0)} (N) = C_F \Big[  \frac{3}{2}  + \frac{1}{N}  - \frac{1}{N+1} 
-2(\gamma_E + \psi(N+1) ) \Big]\,,
\eea 
with $\psi(n) = \Gamma'(n)/\Gamma(n)$ is the di-Gamma function. 

For completeness, we also give the results for gluon induced hard process,
\begin{align}
I_{gg}^{(0)}(N, \theta^2) =&\ \frac{\alpha_s}{2 \pi} \left( \gamma_{gg}^{(0)}(N) -  \gamma_{gg}^{(0)}(N+1) \right) \frac{1}{\theta^2}  \,, 
\\
I_{gq}^{(0)}(N, \theta^2) =&\ \frac{\alpha_s}{2 \pi} \left( \gamma_{gq}^{(0)}(N) -  \gamma_{gq}^{(0)}(N+1) \right) \frac{1}{\theta^2}  \,, 
\end{align}
where
\begin{align}
    \gamma_{gg}^{(0)}(N) =&\ 2 C_A \left[  \frac{1}{N (N-1)} + \frac{1}{(N+1)(N+2)} - (\gamma_E + \psi(N+1) ) \right] +   \frac{11}{6} C_A - \frac{1}{3} n_f  \,,
    \\
    \gamma_{gq}^{(0)}(N) = &\ C_F \frac{2 +N + N^2}{N (N^2 - 1)} \,.
\end{align}

\end{itemize}

\subsection{relation to the TMD PDFs} 
Now we consider the case in which $Q\theta >\Lambda_{\rm QCD}$ but is not significantly larger. Therefore,   
 the incoming gluon transverse momentum $l_t$ could have non-negligible contribution to the $f_{\rm EEC}$ and should be kept in the calculation, leading to potential sensitivities to TMD PDFs. In principle coupling is strong in this region and perturbation theory may not be reliable, nevertheless we can perform a calculation with an effective strong coupling to illustrate the idea. We leave detailed phenomenolgical studies to the future work~\cite{future:2023tap}. Again, we use $P$ to denote the nucleon momentum, $p$ the partonic momentum that entering the hard scattering, $q$ the momentum of detected parton in the forward direction, and $\vec l_t$ intrinsic transverse momentum of the parton from the nucleon.
 
 The calculation essentially follows Eq.~(\ref{eq:split}), but now the gluon transverse momentum is kept to find 
\bea 
f_{q, \rm EEC}(N, \theta) &=& 
 \int dz z^{N-1}
\frac{1}{2}
\int \frac{dq^+ d^2q_t}{q^+(2\pi)^3}  
\frac{q^+}{P} 
\delta((1-z) P^+-q^+ - p_X^+) 
\delta(\theta^2_q - \theta^2)\nn\\
&& \times g_s^2\, {\rm Tr}[T_aT_b]
\frac{1}{2}
{\rm Tr}[\gamma^\mu_t \gamma_{t,\mu} {q {\!\!\!  /}} \gamma^+ ] \left( 
\frac{ {-\vec{q}^2_t} }{(q^+)^2} 
+ \frac{ {-\vec{p}^2_t} }{(p^+)^2}  
\right)
\left(
\frac{p^+}{p^2} \right)^2  \nn \\
&& \times \frac{1}{2} \frac{1}{N_C^2-1}\langle P|
\, A_{t, a,\mu}|X\rangle \langle X| A_t^{a,\mu} |P \rangle  \int d\xi P^+ \delta((1-\xi)P^+-p_X^+)
\int  d^2 l_t \delta^{(2)}(P_t - l_t - p_{X,t})  \,.
\eea 
Since $l_t$ is not negligible, we will have for the single parton emission $\vec p_t = \vec l_t - \vec q_t$, while 
\bea 
p^2 &=& (l-q)^2 = -2l\cdot q = - l^+ q^- 
- l^- q^+ 
+ 2 \vec{l}_t \cdot \vec{q}_t
= - \frac{l^+}{q^+}\vec{q}^2_t 
+ 2 \vec{l}_t \cdot \vec{q}_t
- \frac{q^+}{l^+} \vec{l}_t^2 \nn \\ 
&=& -\frac{l^+}{q^+} \left(\vec{q}_t - \frac{q^+}{l^+}\vec{l}_t \right)^2 \equiv 
-\frac{l^+}{q^+} \vec{K}_t^2 \,, 
\eea 
where $q^+ = \xi (1 - z/\xi) P^+$ and $l^+ = \xi P^+$.
Therefore we find 
\bea 
f_{q, \rm EEC}(N, \theta) &=& 
\int \, 
\frac{d\xi}{\xi}  
\int dz z^{N-1}
\left(1-\frac{z}{\xi} \right)\xi 
\nn\\
&& \times \frac{\alpha_s}{2\pi^2} T_R 
\frac{ 1}{ 2\theta^2} 
\int  d\phi d^2 l_t   
 \, 
  q_t^2 \left[ \vec{q}_t^2 \left( \frac{z}{\xi} \right)^2
+ (\vec{l}_t-\vec{q}_t)^2 \left(1-\frac{z}{\xi}\right)^2
\right]
\left(
\frac{ 1}{  \vec{K}_t^2 } \right)^2\nn \\
&& \times \sum_X (-)\langle P|
\, A_{t, a,\mu}|X\rangle \langle X| A_t^{a,\mu} |P \rangle   \xi P^+ \delta((1-\xi)P^+-p_X^+)
  \delta^{(2)}(P_t - l_t - p_{X,t}) \,, 
\eea 
where we have worked out the $\vec{q}_t$ integration by noting that $\vec{q}_t = \left(1-\frac{z}{\xi}\right)\xi \frac{P^+}{2}\theta(\cos\phi,\sin\phi)$.
Now we note that the last line is related to the unpolarized gluon TMD PDF. Using a change of variable $z/\xi \to x$, we find
\bea 
f_{q, \rm EEC}(N, \theta) &=& 
 \int   d^2 l_t  
\left[ \frac{\alpha_s}{2\pi} T_R 
\frac{ 1}{ \theta^2} 
 \,  \int dx\, x^{N-1}(1-x)  \int \frac{d\phi}{2\pi}
  q_t^2 \left[ \vec{q}_t^2 x^2
+ (\vec{l}_t-\vec{q}_t)^2 \left(1- x \right)^2
\right]
\left(
\frac{ 1}{  \vec{K}_t^2 } \right)^2 \right] \times  \int d\xi \xi^{N} f_{g/P}(\xi,\vec{l}_t)
\nn
\\
&=& \int d^2 l_t \, I^{(0)}(N, \theta^2, \vec{l}_t) f_{g/P}(N+1, \vec{l}_t) \,,
 \label{eq:tmd_matching}
\eea 
where $\vec{K}_t^2 = (\vec{q}_t - (1 - x ) \vec{l}_t)^2$.
Note that this result reduces exactly to the collinear case in eq.~(\ref{eq:coll}) if we let $\vec{l}_t = 0$, as it should.
Eq.~(\ref{eq:tmd_matching}) suggests that the nucleon EEC probes the moment of the TMD PDFs with the matching coefficient $I^{(0)}(N,\theta^2,\vec{l}_t)$, which differs from the matching coefficient in the last section in that it depends on $\vec l_t$.  Again, we note that due to the overall $(1-x)$ factor, there is no perturbative Sudakov logarithms in the matching coefficient. 

\end{widetext} 

\bibliographystyle{h-physrev}   
\bibliography{refs}

\begin{thebibliography}{10}

\bibitem{AbdulKhalek:2021gbh}
R.~Abdul~Khalek {\em et~al.},
\newblock (2021), 2103.05419.

\bibitem{Proceedings:2020eah}
{\em {Proceedings, Probing Nucleons and Nuclei in High Energy Collisions:
  Dedicated to the Physics of the Electron Ion Collider}: {Seattle (WA), United
  States, October 1 - November 16, 2018}}, WSP, 2020, 2002.12333.

\bibitem{Anderle:2021wcy}
D.~P. Anderle {\em et~al.},
\newblock Front. Phys. (Beijing) {\bf 16}, 64701 (2021), 2102.09222.

\bibitem{Collins:2005ie}
J.~C. Collins {\em et~al.},
\newblock Phys. Rev. D {\bf 73}, 014021 (2006), hep-ph/0509076.

\bibitem{Vogelsang:2005cs}
W.~Vogelsang and F.~Yuan,
\newblock Phys. Rev. D {\bf 72}, 054028 (2005), hep-ph/0507266.

\bibitem{HERMES:2009lmz}
HERMES, A.~Airapetian {\em et~al.},
\newblock Phys. Rev. Lett. {\bf 103}, 152002 (2009), 0906.3918.

\bibitem{Bacchetta:2011gx}
A.~Bacchetta and M.~Radici,
\newblock Phys. Rev. Lett. {\bf 107}, 212001 (2011), 1107.5755.

\bibitem{Echevarria:2014xaa}
M.~G. Echevarria, A.~Idilbi, Z.-B. Kang, and I.~Vitev,
\newblock Phys. Rev. D {\bf 89}, 074013 (2014), 1401.5078.

\bibitem{Scimemi:2019cmh}
I.~Scimemi and A.~Vladimirov,
\newblock JHEP {\bf 06}, 137 (2020), 1912.06532.

\bibitem{Gutierrez-Reyes:2018qez}
D.~Gutierrez-Reyes, I.~Scimemi, W.~J. Waalewijn, and L.~Zoppi,
\newblock Phys. Rev. Lett. {\bf 121}, 162001 (2018), 1807.07573.

\bibitem{Liu:2018trl}
X.~Liu, F.~Ringer, W.~Vogelsang, and F.~Yuan,
\newblock Phys. Rev. Lett. {\bf 122}, 192003 (2019), 1812.08077.

\bibitem{Gutierrez-Reyes:2019msa}
D.~Gutierrez-Reyes, Y.~Makris, V.~Vaidya, I.~Scimemi, and L.~Zoppi,
\newblock JHEP {\bf 08}, 161 (2019), 1907.05896.

\bibitem{Gutierrez-Reyes:2019vbx}
D.~Gutierrez-Reyes, I.~Scimemi, W.~J. Waalewijn, and L.~Zoppi,
\newblock JHEP {\bf 10}, 031 (2019), 1904.04259.

\bibitem{Arratia:2020nxw}
M.~Arratia, Z.-B. Kang, A.~Prokudin, and F.~Ringer,
\newblock Phys. Rev. D {\bf 102}, 074015 (2020), 2007.07281.

\bibitem{Liu:2020dct}
X.~Liu, F.~Ringer, W.~Vogelsang, and F.~Yuan,
\newblock Phys. Rev. D {\bf 102}, 094022 (2020), 2007.12866.

\bibitem{Arratia:2020ssx}
M.~Arratia, Y.~Makris, D.~Neill, F.~Ringer, and N.~Sato,
\newblock Phys. Rev. D {\bf 104}, 034005 (2021), 2006.10751.

\bibitem{Li:2020rqj}
H.~T. Li and I.~Vitev,
\newblock Phys. Rev. Lett. {\bf 126}, 252001 (2021), 2010.05912.

\bibitem{Kang:2020fka}
Z.-B. Kang, X.~Liu, S.~Mantry, and D.~Y. Shao,
\newblock Phys. Rev. Lett. {\bf 125}, 242003 (2020), 2008.00655.

\bibitem{H1:2021wkz}
H1, V.~Andreev {\em et~al.},
\newblock Phys. Rev. Lett. {\bf 128}, 132002 (2022), 2108.12376.

\bibitem{Kang:2021kpt}
Z.-B. Kang, J.~Terry, A.~Vossen, Q.~Xu, and J.~Zhang,
\newblock Phys. Rev. D {\bf 105}, 094033 (2022), 2108.05383.

\bibitem{Liu:2021ewb}
X.~Liu and H.~Xing,
\newblock (2021), 2104.03328.

\bibitem{Kang:2021ffh}
Z.-B. Kang, K.~Lee, D.~Y. Shao, and F.~Zhao,
\newblock JHEP {\bf 11}, 005 (2021), 2106.15624.

\bibitem{Li:2021gjw}
H.~T. Li, Z.~L. Liu, and I.~Vitev,
\newblock Phys. Lett. B {\bf 827}, 137007 (2022), 2108.07809.

\bibitem{Lai:2022aly}
W.~K. Lai, X.~Liu, M.~Wang, and H.~Xing,
\newblock (2022), 2205.04570.

\bibitem{Kang:2022dpx}
Z.-B. Kang, K.~Lee, D.~Y. Shao, and F.~Zhao,
\newblock (2022), 2201.04582.

\bibitem{Basham:1978bw}
C.~Basham, L.~S. Brown, S.~D. Ellis, and S.~T. Love,
\newblock Phys. Rev. Lett. {\bf 41}, 1585 (1978).

\bibitem{Basham:1978zq}
C.~Basham, L.~Brown, S.~Ellis, and S.~Love,
\newblock Phys. Rev. D {\bf 19}, 2018 (1979).

\bibitem{Chen:2020vvp}
H.~Chen, I.~Moult, X.~Zhang, and H.~X. Zhu,
\newblock Phys. Rev. D {\bf 102}, 054012 (2020), 2004.11381.

\bibitem{Hofman:2008ar}
D.~M. Hofman and J.~Maldacena,
\newblock JHEP {\bf 05}, 012 (2008), 0803.1467.

\bibitem{Belitsky:2013ofa}
A.~Belitsky, S.~Hohenegger, G.~Korchemsky, E.~Sokatchev, and A.~Zhiboedov,
\newblock Phys. Rev. Lett. {\bf 112}, 071601 (2014), 1311.6800.

\bibitem{Belitsky:2013xxa}
A.~Belitsky, S.~Hohenegger, G.~Korchemsky, E.~Sokatchev, and A.~Zhiboedov,
\newblock Nucl. Phys. B {\bf 884}, 305 (2014), 1309.0769.

\bibitem{Kologlu:2019mfz}
M.~Kologlu, P.~Kravchuk, D.~Simmons-Duffin, and A.~Zhiboedov,
\newblock JHEP {\bf 01}, 128 (2021), 1905.01311.

\bibitem{Korchemsky:2019nzm}
G.~Korchemsky,
\newblock JHEP {\bf 01}, 008 (2020), 1905.01444.

\bibitem{Dixon:2019uzg}
L.~J. Dixon, I.~Moult, and H.~X. Zhu,
\newblock Phys. Rev. D {\bf 100}, 014009 (2019), 1905.01310.

\bibitem{Chen:2019bpb}
H.~Chen {\em et~al.},
\newblock JHEP {\bf 08}, 028 (2020), 1912.11050.

\bibitem{Chen:2020adz}
H.~Chen, I.~Moult, and H.~X. Zhu,
\newblock Phys. Rev. Lett. {\bf 126}, 112003 (2021), 2011.02492.

\bibitem{Chang:2020qpj}
C.-H. Chang, M.~Kologlu, P.~Kravchuk, D.~Simmons-Duffin, and A.~Zhiboedov,
\newblock JHEP {\bf 05}, 059 (2022), 2010.04726.

\bibitem{Li:2021zcf}
Y.~Li, I.~Moult, S.~S. van Velzen, W.~J. Waalewijn, and H.~X. Zhu,
\newblock Phys. Rev. Lett. {\bf 128}, 182001 (2022), 2108.01674.

\bibitem{Jaarsma:2022kdd}
M.~Jaarsma, Y.~Li, I.~Moult, W.~Waalewijn, and H.~X. Zhu,
\newblock JHEP {\bf 06}, 139 (2022), 2201.05166.

\bibitem{Komiske:2022enw}
P.~T. Komiske, I.~Moult, J.~Thaler, and H.~X. Zhu,
\newblock (2022), 2201.07800.

\bibitem{Holguin:2022epo}
J.~Holguin, I.~Moult, A.~Pathak, and M.~Procura,
\newblock (2022), 2201.08393.

\bibitem{Yan:2022cye}
K.~Yan and X.~Zhang,
\newblock Phys. Rev. Lett. {\bf 129}, 021602 (2022), 2203.04349.

\bibitem{Chen:2022jhb}
H.~Chen, I.~Moult, J.~Sandor, and H.~X. Zhu,
\newblock (2022), 2202.04085.

\bibitem{Chang:2022ryc}
C.-H. Chang and D.~Simmons-Duffin,
\newblock (2022), 2202.04090.

\bibitem{Chen:2022swd}
H.~Chen, I.~Moult, J.~Thaler, and H.~X. Zhu,
\newblock JHEP {\bf 07}, 146 (2022), 2205.02857.

\bibitem{Lee:2022ige}
K.~Lee, B.~Me\c{c}aj, and I.~Moult,
\newblock (2022), 2205.03414.

\bibitem{Larkoski:2022qlf}
A.~J. Larkoski,
\newblock (2022), 2205.12375.

\bibitem{Ricci:2022htc}
L.~Ricci and M.~Riembau,
\newblock (2022), 2207.03511.

\bibitem{Yang:2022tgm}
T.-Z. Yang and X.~Zhang,
\newblock (2022), 2208.01051.

\bibitem{Sveshnikov:1995vi}
N.~A. Sveshnikov and F.~V. Tkachov,
\newblock Phys. Lett. B {\bf 382}, 403 (1996), hep-ph/9512370.

\bibitem{Tkachov:1995kk}
F.~V. Tkachov,
\newblock Int. J. Mod. Phys. A {\bf 12}, 5411 (1997), hep-ph/9601308.

\bibitem{Korchemsky:1999kt}
G.~P. Korchemsky and G.~F. Sterman,
\newblock Nucl. Phys. B {\bf 555}, 335 (1999), hep-ph/9902341.

\bibitem{Bauer:2008dt}
C.~W. Bauer, S.~P. Fleming, C.~Lee, and G.~F. Sterman,
\newblock Phys. Rev. D {\bf 78}, 034027 (2008), 0801.4569.

\bibitem{Bauer:2000yr}
C.~W. Bauer, S.~Fleming, D.~Pirjol, and I.~W. Stewart,
\newblock Phys. Rev. D {\bf 63}, 114020 (2001), hep-ph/0011336.

\bibitem{Bauer:2001yt}
C.~W. Bauer, D.~Pirjol, and I.~W. Stewart,
\newblock Phys. Rev. D {\bf 65}, 054022 (2002), hep-ph/0109045.

\bibitem{Bauer:2001ct}
C.~W. Bauer and I.~W. Stewart,
\newblock Phys. Lett. B {\bf 516}, 134 (2001), hep-ph/0107001.

\bibitem{Beneke:2002ph}
M.~Beneke, A.~P. Chapovsky, M.~Diehl, and T.~Feldmann,
\newblock Nucl. Phys. B {\bf 643}, 431 (2002), hep-ph/0206152.

\bibitem{Bauer:2002nz}
C.~W. Bauer, S.~Fleming, D.~Pirjol, I.~Z. Rothstein, and I.~W. Stewart,
\newblock Phys. Rev. D {\bf 66}, 014017 (2002), hep-ph/0202088.

\bibitem{Li:2020bub}
H.~T. Li, I.~Vitev, and Y.~J. Zhu,
\newblock JHEP {\bf 11}, 051 (2020), 2006.02437.

\bibitem{Ali:2020ksn}
A.~Ali, G.~Li, W.~Wang, and Z.-P. Xing,
\newblock Eur. Phys. J. C {\bf 80}, 1096 (2020), 2008.00271.

\bibitem{Li:2021txc}
H.~T. Li, Y.~Makris, and I.~Vitev,
\newblock Phys. Rev. D {\bf 103}, 094005 (2021), 2102.05669.

\bibitem{Arratia:2022quz}
M.~Arratia {\em et~al.},
\newblock (2022), 2208.05472.

\bibitem{Cebra:2022avc}
D.~Cebra, X.~Dong, Y.~Ji, S.~R. Klein, and Z.~Sweger,
\newblock (2022), 2204.07915.

\bibitem{Bylinkin:2022rxd}
A.~Bylinkin {\em et~al.},
\newblock (2022), 2208.14575.

\bibitem{Sjostrand:2014zea}
T.~Sj\"ostrand {\em et~al.},
\newblock Comput. Phys. Commun. {\bf 191}, 159 (2015), 1410.3012.

\bibitem{Sivers:1989cc}
D.~W. Sivers,
\newblock Phys. Rev. D {\bf 41}, 83 (1990).

\bibitem{Collins:2002kn}
J.~C. Collins,
\newblock Phys. Lett. B {\bf 536}, 43 (2002), hep-ph/0204004.

\bibitem{future:2023tap}
H.~Cao, X.~Liu, and H.~X. Zhu.

\bibitem{Stewart:2010qs}
I.~W. Stewart, F.~J. Tackmann, and W.~J. Waalewijn,
\newblock JHEP {\bf 09}, 005 (2010), 1002.2213.

\end{thebibliography}

\end{document}